\documentclass[hidelinks, conference,a4paper]{IEEEtran}
\IEEEoverridecommandlockouts

\usepackage{cite}
\usepackage{amsmath,amssymb,amsfonts}
\usepackage{algorithmic}
\usepackage{graphicx}
\usepackage{textcomp}
\usepackage{booktabs}
\usepackage{siunitx}
\usepackage{xcolor}
\usepackage{subcaption}
\usepackage[nolist]{acronym}

\def\BibTeX{{\rm B\kern-.05em{\sc i\kern-.025em b}\kern-.08em
    T\kern-.1667em\lower.7ex\hbox{E}\kern-.125emX}}
\begin{document}

\begin{acronym}
	\acro{acMFCC}[MFCC]{Mel Frequency Cepstral Coefficients}
	\acro{acIMFCC}[IMFCC]{Inverse Mel Frequency Cepstral Coefficients}
	\acro{acLFCC}[LFCC]{Linear Frequency Cepstral Coefficients}
	\acro{acPCA}[PCA]{Principal Component Analysis}
\end{acronym}

\title{An Acoustical Machine Learning Approach to Determine Abrasive Belt Wear of Wide Belt Sanders\\
	\thanks{This study is partially supported by the European Social Funds (ESF) \mbox{No. R.6-V0332.2.43/1/5}.}
}

\author{
	\IEEEauthorblockN{
		Maximilian Bundscherer\,\IEEEauthorrefmark{1}\IEEEauthorrefmark{2},
		Thomas H. Schmitt\,\IEEEauthorrefmark{2},
		Sebastian Bayerl\,\IEEEauthorrefmark{2},
		Thomas Auerbach\IEEEauthorrefmark{3} and
		Tobias Bocklet\IEEEauthorrefmark{2},
	}
	\IEEEauthorblockA{\IEEEauthorrefmark{2}\textit{Department of Computer Science} - \textit{Technische Hochschule Nürnberg Georg Simon Ohm} Nürnberg, Germany}
	\IEEEauthorblockA{\IEEEauthorrefmark{3}\textit{Production Technology} - \textit{Hans Weber Maschinenfabrik GmbH} Kronach, Germany}
	\IEEEauthorblockA{\IEEEauthorrefmark{1}Email: maximilian.bundscherer@th-nuernberg.de}
}

\maketitle

\begin{abstract}

This paper describes a machine learning approach to determine the abrasive belt wear of wide belt sanders used in industrial processes based on acoustic data, regardless of the sanding process-related parameters, Feed speed, Grit Size, and Type of material.
Our approach utilizes Decision Tree, Random Forest, k-nearest Neighbors, and Neural network Classifiers to detect the belt wear from Spectrograms, Mel Spectrograms, MFCC, IMFCC, and LFCC, yielding an accuracy of up to 86.1\% on five levels of belt wear.
A 96\% accuracy could be achieved with different Decision Tree Classifiers specialized in different sanding parameter configurations.
The classifiers could also determine with an accuracy of 97\% if the machine is currently sanding or is idle and with an accuracy of 98.4\% and 98.8\% detect the sanding parameters Feed speed and Grit Size.
We can show that low-dimensional mappings of high-dimensional features can be used to visualize belt wear and sanding parameters meaningfully.

\end{abstract}


\section{Introduction}

\begin{figure}
	\centerline{\includegraphics[width=0.99\linewidth]{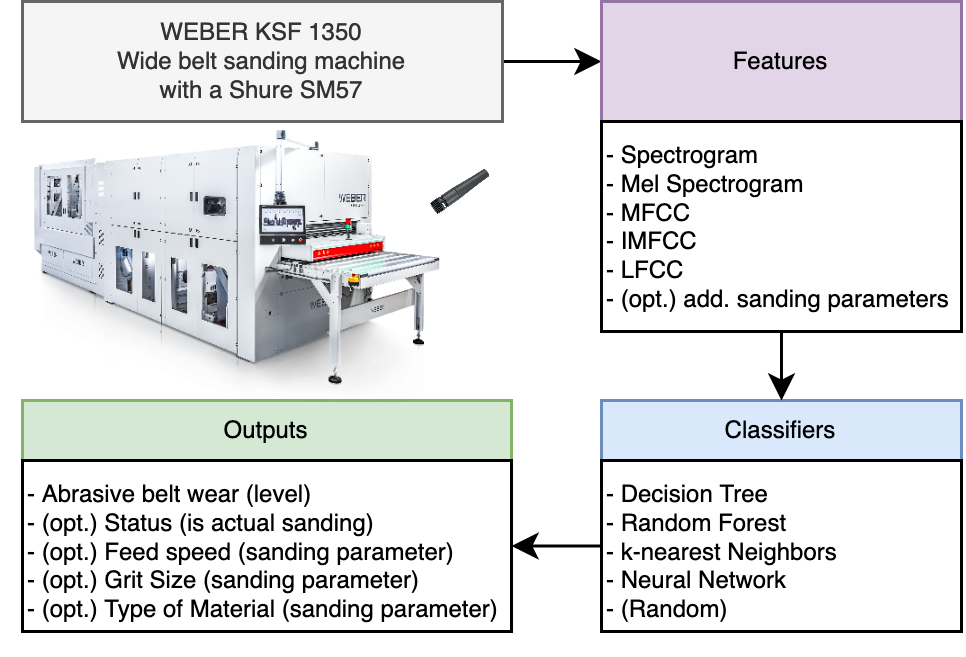}}
	\caption{
		Schematic overview of our approach.
	}
	\label{figStructure}
\end{figure}

Wide belt sanding machines are commonly used in industrial processes to remove material from a workpiece by an abrasive belt \cite{GrindCurentSound}.
This paper uses acoustic data and machine learning methods to determine the abrasive belt wear of a WEBER KSF 1350 wide belt sander.
Our models aim to optimize tool change intervals, a critical optimization criterion in industrial processes\cite{ToolWearOpti}.
We used lightweight approaches that potentially allow our models to be deployed on low-cost endpoints such as a Raspberry Pi or ESP32.
The use of acoustic sensors, like microphones as additional sensors and low-cost edge devices for classification in industrial processes, is convenient and inexpensive since no machines or production chains need to be strongly modified or rebuilt.
Fig. \ref{figStructure} gives a schematic overview of our approach.

\subsection{Our contributions}

\begin{itemize}
	\item Recording and labeling of wide belt sanding machine operations with 18 sanding process-related parameter configurations.
	\item Describing and evaluating a machine learning based method for classifying five levels of belt wear, machine state (is actual sanding or is idle), and sanding parameters.
	\item Meaningful low-dimensional mapping of high-dimensional features.
	\item Models that use acoustic data only and can operate on lightweight edge devices.
\end{itemize}

\subsection{Related work}

In the literature, a distinction is made between the terms grinding, sanding, and polishing to describe abrasive processes.
\cite{GrindCurentSound} presents an approach to detect abrasive grinding belt wear over three levels on a grinding machine using sound and current as features under varying grinding parameters. 
Convolutional Neural Networks are used to predict belt grinding tool wear in a polishing process using 3-axes force and vibration data as models input by \cite{GrinderCNNPred}.
Other publications have focused on details of the grinding process \cite{GrinderDetail1} and tool condition monitoring \cite{GrinderDetail2}.
This study's novelty is applying acoustic machine learning detection to wide belt sanding machines and focuses on the varying wood sanding parameters Feed speed, Grit Size, and Type of material.

\section{Data}
\label{lblSectionData}

Wide belt sanding machines have sanding process-related parameters that influence the acoustics characteristics of the process more than the abrasive belt wear itself.
This phenomenon has also been observed in grinding machines \cite{GrinderSoundParam}.
The sander was allowed to cool down between recordings to mitigate the influence of different engine temperatures.

\subsection{Sanding Parameters}
\label{lblSubSandingParams}
\sisetup{per-mode=symbol}
We recorded sanding operations with three different Feed speeds
$\in \{10, 14, 18\}$ ($m/min$), three different Grit Sizes
$\in \{80, 150, 240\}$, and two Types of materials
$\in \{\text{soft}, \text{hard}\}$.
Respectively, 18 sanding parameter configurations were tested to detect the belt wear regardless of these parameters.
Three workpiece positions $\in \{\text{Left}, \text{Center}, \text{Right}\}$, related to the sanding belt, were considered.
Each sanding operation was recorded threefold to ensure robustness to noise.
Considering five levels of belt wear, we recorded a total of 810 sanding operations.

\subsection{Recording Setup}
\label{lblSubRecordingSetup}

The used microphone, Shure SM57, hung near the sanding belt and was attached to a Focusrite Scarlett 2i2 interface.
The sanding sounds were recorded with a sampling rate of %
${\SI{44.1}{\kilo\hertz}}$
According to the Nyquist-Shannon sampling theorem \cite{SAMPLINGThero}, with this choice, it is possible to save maximum frequencies up to
${\frac{\SI{44.1}{\kilo\hertz}}{2} = \SI{22.05}{\kilo\hertz}}$.
The Shure SM57 microphone does not pick up frequencies linearly. Its recording quality deteriorates increasingly above
${\SI{10}{\kilo\hertz}}$
\cite{ShureSM57}.

\subsection{Pre-selection of sanding}
\label{lblSubSectionPreSelection}

\begin{figure}
	\centerline{\includegraphics[width=0.99\linewidth]{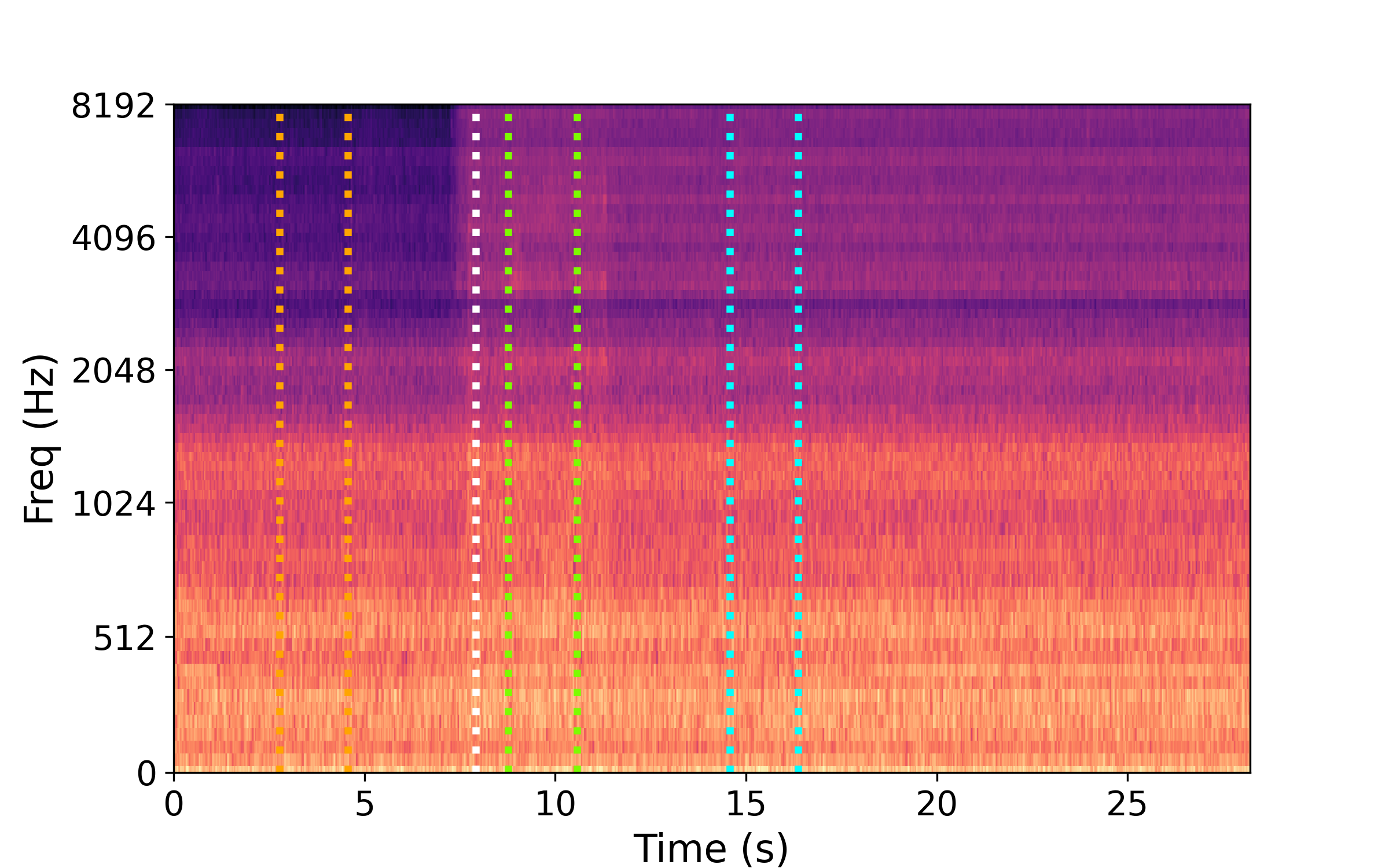}}
	\caption{Mel Spectrogram visualization of a sanding process with (orange) pre-sanding, (green) actual-sanding, (cyan) post-sanding time markers, and (white) Mel Spectrogram energy detected start-marker.}
	\label{figPreselect}
\end{figure}

The recordings have a duration of 2 minutes. The actual sanding itself only takes approximately 3-4 seconds.
Process-related noises such as a "flop" noise at the beginning of the actual sanding process and the increasingly quieter sanding noise at the end \cite{GrindSoundProc} should not be considered.
All recordings were cut down to 1.8 seconds by adding a fixed offset of 1.3 seconds to the actual start-marker and setting the stop-marker with a fixed duration of 1.8 seconds related to the start-marker.
Possible sanding operations were pre-selected using a Mel Spectrogram-based energy detection to make pre-selection more efficient.
To increase the label's credibility, two people did the sanding pre-selection.
A classifier that can determine if the machine is currently sanding or idle, described later in the section \ref{lblSectionMethods}, was used to identify and adjust faulty selections.
This classifier requires sound samples before actual sanding has been taken (pre-sanding), where only the sanding engine is hearable, and after sanding has been taken (post-sanding), where the engine and the dust collector are hearable.
These markers were set at a fixed duration from the actual start- and stop-markers, shown in Fig. \ref{figPreselect}.

\subsection{Abrasive belt wear labeling}
\label{lblSubSectionWearLabeling}

The abrasive belt wear is a progressive process that can be described with a continuous value, e.g. as a percentage value.
Precisely determining wear is problematic because, among other things, the abrasive belt does not always wear out the same\cite{GrindMatNotLinearRemove}.
The expert who labelled the belt wear specified the abrasive belt wear as five discrete values:
0 (fresh), 1 (0-25\%  worn), 2 (25-50\% worn), 3 (50-75\% worn), and 4 (75-100\% worn).

\subsection{Feature Extraction}
\label{lblSubSectionFeatureExtraction}

In our experiments, we used Spectrograms, Mel Spectrograms, \ac{acMFCC}, \ac{acIMFCC}, and \ac{acLFCC} as feature extraction methods, as shown in Fig. \ref{figStructure}.
The Mel scale emulates hearing auditory perception, with a higher frequency resolution in the lower frequencies and a coarser resolution in the higher frequencies \cite{MelSCale}.
Mel Spectrograms and \ac{acMFCC} are extracted with the help of Mel, influenced by this behaviour \cite{CiteMFCC}.
\ac{acIMFCC} uses inverted Mel filters with a higher frequency resolution in the higher frequencies and a coarser resolution in the lower frequencies\cite{CiteIMFCC}.
\ac{acLFCC} use a linear filter instead of a Mel filter and are commonly used in high-frequency domains \cite{CiteLFCC}.
Hyperparameters for spectral features where chosen via extensive grid search \cite{CiteGS} from parameters: 
\begin{itemize}
	\item ${w_{l}} \in \{32, 64, 128\}$ (windowed signal length $ms$)
	\item ${n_{b}} \in \{32, 64, 128\}$ (number of mel/linear bins)
	\item ${n_{c}} \in \{20, 40, 60\}$ (number of mel/linear coefficients)
\end{itemize}
The ${h_{l}}$ (hop length) was modelled as a function of the ${w_{l}}$, with
${h_{l} = \frac{1}{4} w_{l}}$
.
The used window function is hann.
The features were mapped to a 1D vector by concatenation, since there are no significant changes within a single sanding operation.

\section{Methods}
\label{lblSectionMethods}

To train and evaluate our supervised classifiers, we used the discretized labels from the expert, see section \ref{lblSubSectionWearLabeling}.
We utilize Decision Tree, Random Forest, k-nearest Neighbors, Neural network, and Random Classifiers as representative of the supervised classifiers to detect the abrasive belt wear, as shown in Fig. \ref{figStructure}.
The final classifier is based on class distribution and is mainly used to see if the models are better than guessing \cite{MeasClassi}.
To achieve better results in tool wear detection, we tested adding sanding process-related parameters to the model's features. 
In addition, we also try to monitor with these classifiers if the machine is currently actively sanding or idling and also to determine the sanding parameters themselves.
Accuracy scores and confusion matrices were used to evaluate these classifiers.
Data are split into disjunct train and test sets, using the first and second repetitions of the sanding process as the training set and the third and last repetitions as the test set.
Unsupervised \ac{acPCA}-transformed data in visualizations are coloured according to their properties, such as the abrasive belt wear or the sanding parameters.

\section{Results}
\label{lblSectionResults}

The feature extraction hyperparameters, see section \ref{lblSubSectionFeatureExtraction}, we tested did not significantly affect the average classifier's accuracies (about 0.2-1\%) and \ac{acPCA} transformations, so we focus on the best fitting values of the extensive grid search:
${w_{l}} = \SI{64}{\milli\second}$, ${n_{b} = 64}$ and ${n_{c} = 40}$
in this paper.
The proper selection of the feature extraction method and the classifier used are much more critical to the accuracy achieved.

\subsection{Supervised Classifiers}

\begin{table}
	\begin{center}
		\caption{Accuracies of Decision Tree (Dec Tree), Random Forest (R Forest), k-nearest Neighbors (KNN), Neural network (NN), and Random Classifiers by feature extraction methods, across all sanding operations, including all parameters.}%
		\label{tabResults}
		\begin{tabular}{@{}llllll@{}}
			\toprule
			& Dec Tree & R Forest & KNN & NN & Random \\ \midrule
			Spectrogram        & 52.9\%              & 49\%                   & 58.7\%    & 40\%    & 20\%           \\
			Mel Spectrogram    & 86.1\%              & 83.2\%                   & 70.7\%     & 84.2\%   & 22\%         \\
			MFCC              & 76.1\%              & 74.5\%                   & 79.6\%       & 74.6\%   & 18\%       \\
			IMFCC            & 55.2\%              & 54.9\%                   & 48.3\%   & 55.5\%   & 19.5\%           \\
			LFCC            & 70.5\%              & 68.8\%                   & 71.5\%     & 73.6\%   & 21\%         \\ \bottomrule
		\end{tabular}
		
	\end{center}
\end{table}

\begin{figure}
	\centering
	\begin{minipage}{0.99\linewidth}
			\centering
			\begin{subfigure}{0.32\linewidth}
					\includegraphics[height=2.9cm, trim=0cm 0cm 0cm 0cm, clip]{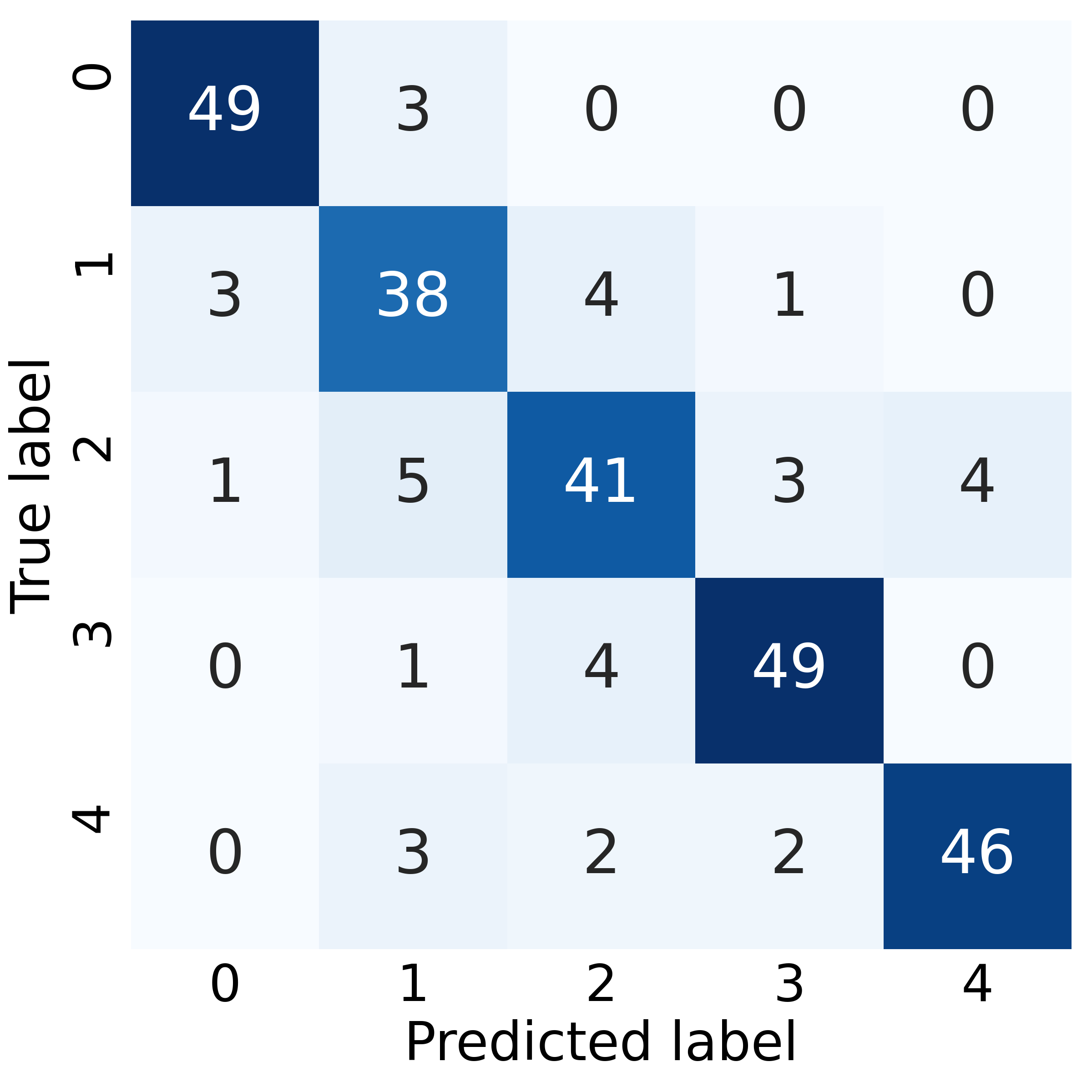}
					\caption{}
					\label{figConAbn}
				\end{subfigure}
			\hfill
			\begin{subfigure}{0.32\linewidth}
					\includegraphics[height=2.9cm, trim=1cm 0cm 0cm 0cm, clip]{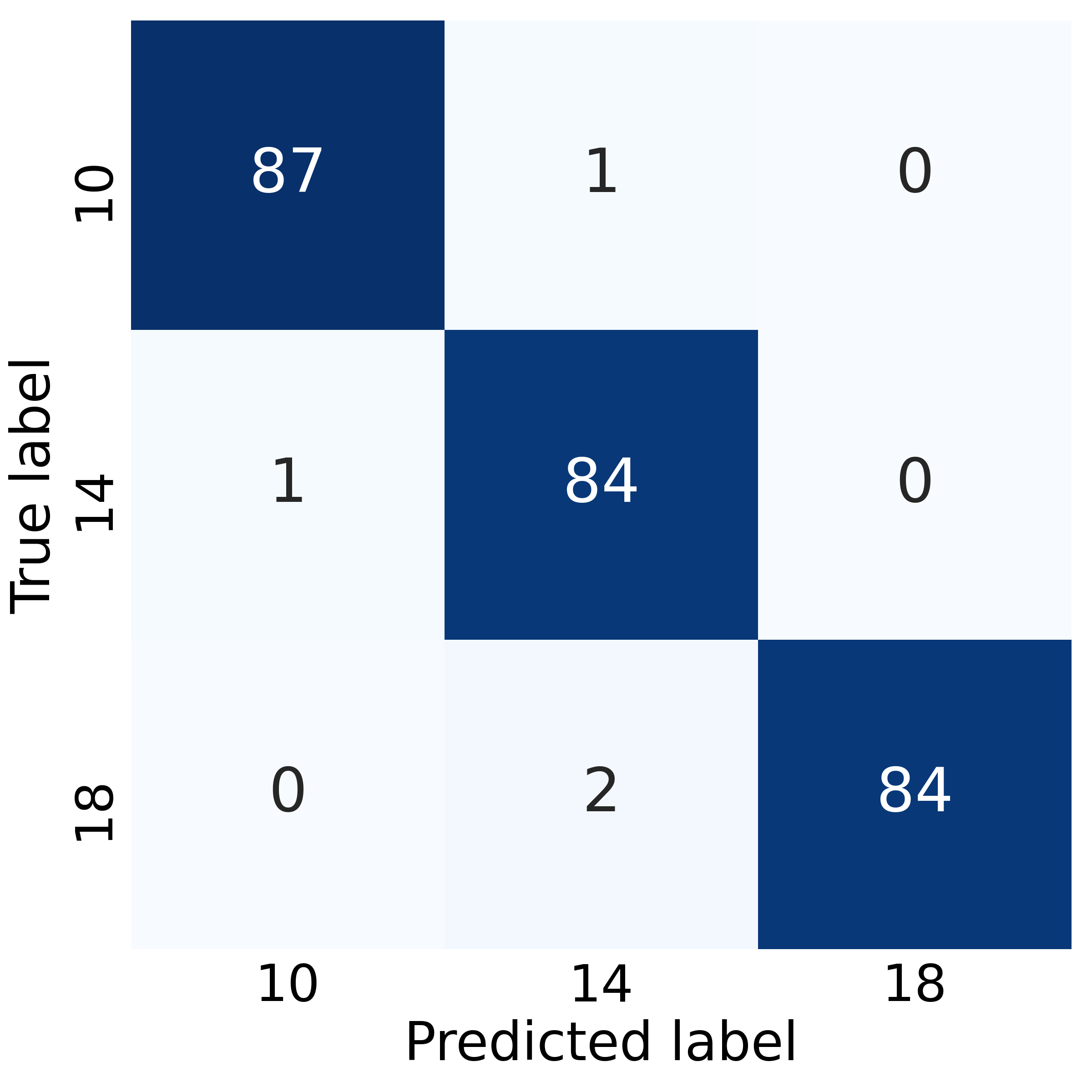}
					\caption{}
					\label{figConSchnitt}
				\end{subfigure}
		\hfill
			\begin{subfigure}{0.32\linewidth}
					\includegraphics[height=2.9cm, trim=1cm 0cm 0cm 0cm, clip]{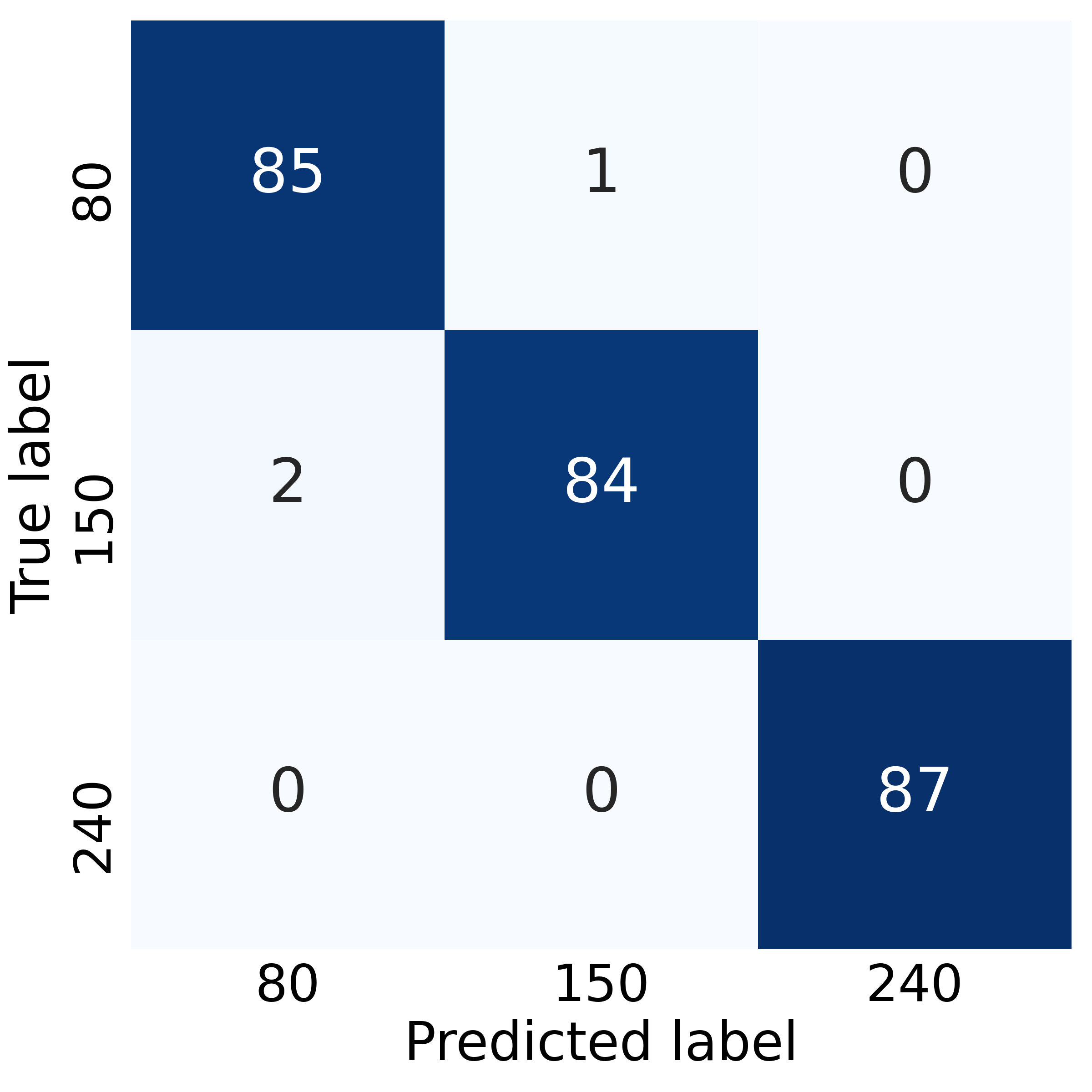}
					\caption{}
					\label{figConKoer}
				\end{subfigure}
			\caption{Confusion matrices of the Decision Tree Classifiers of (a) Abrasive belt wear, (b) Feed speed, and (c) Grit Size.}
			%\label{fig3}
		\end{minipage}
\end{figure}

Decision Tree Classifiers using Mel Spectrograms as input features performs best in detecting the five levels of abrasive belt wear with an 86.1\% accuracy generalized overall tested sanding parameters configurations, as shown in Table \ref{tabResults}.
The confusion matrix for this classifier is shown in Fig. \ref{figConAbn}.
The additional passing of sanding process-related parameters as features to the classifiers does not significantly improve the average accuracy of the classifiers by 0.5\%.
A 96\% average accuracy could be achieved with different Decision Tree Classifiers specialized in different sanding parameter configurations.
Other tested Decision Trees could determine with an accuracy of 97\% if the machine is currently sanding or is idle and with an accuracy of 98.4\% and 98.8\% detect the sanding parameters Feed speed and Grit Size.
Confusion matrices are presented in Fig. \ref{figConSchnitt} and \ref{figConKoer}.
With an accuracy of 51\%, the sanding parameter Type of material could not be reliably detected.

\subsection{Unsupervised Visualizations}

\begin{figure}
	\centering
	\begin{minipage}{0.99\linewidth}
		\centering
		\begin{subfigure}{0.32\linewidth}
			\includegraphics[width=\linewidth, trim=2.0cm 2.0cm 2.0cm 2.0cm, clip]{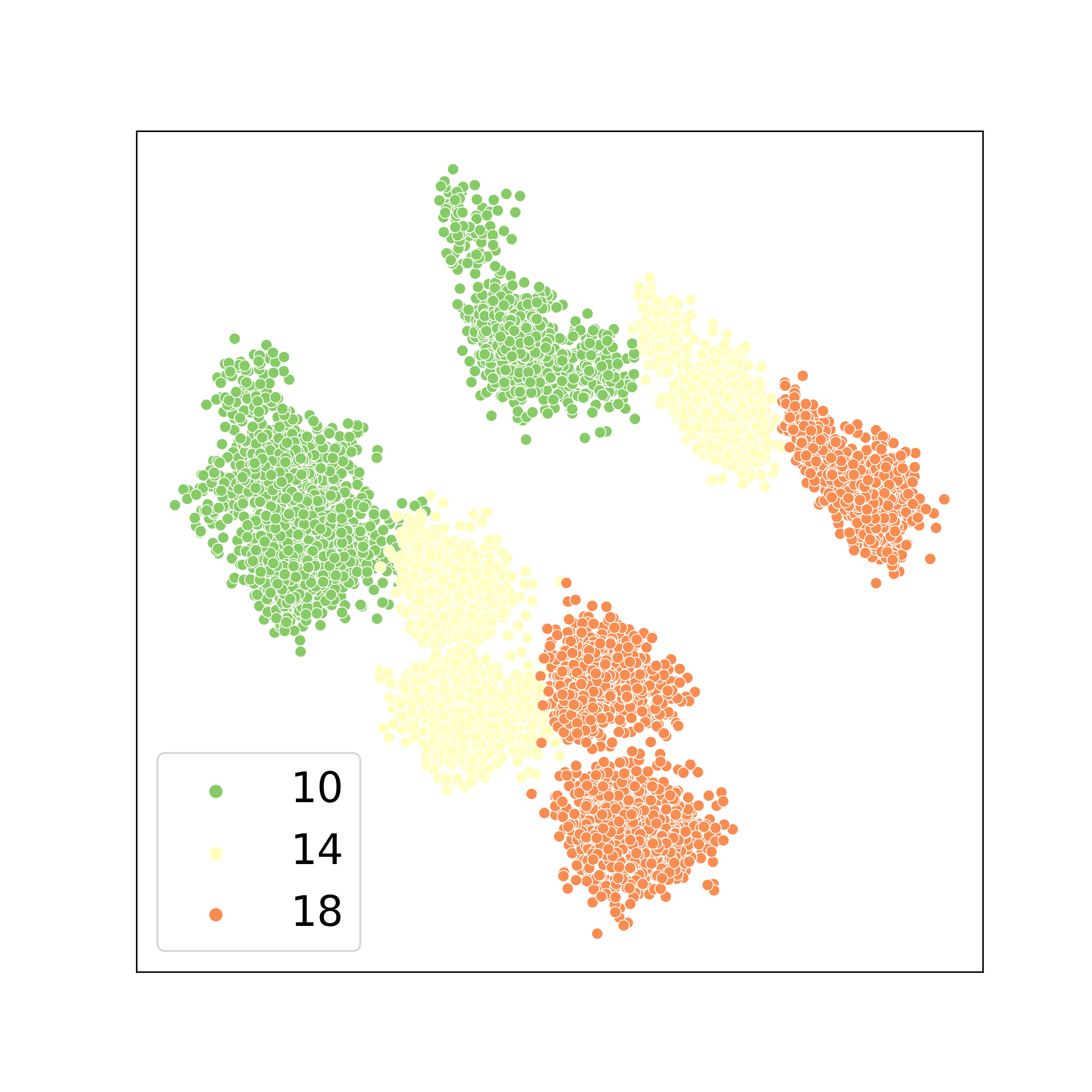}
			\caption{}
			\label{figPCASchnitt}
		\end{subfigure}
		\begin{subfigure}{0.32\linewidth}
			\includegraphics[width=\linewidth, trim=2.0cm 2.0cm 2.0cm 2.0cm, clip]{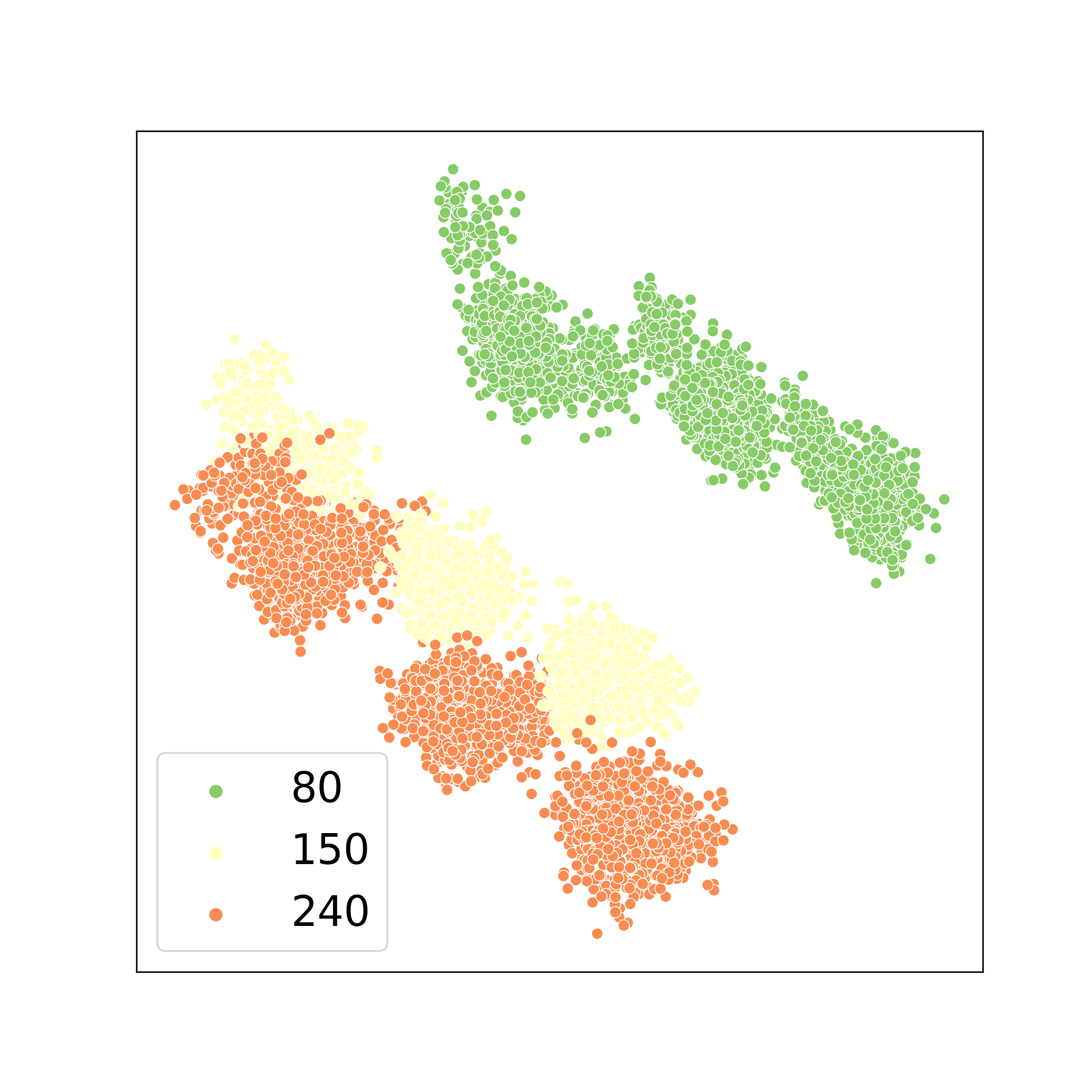}
			\caption{}
			\label{figPCAKoer}
		\end{subfigure}
		\begin{subfigure}{0.32\linewidth}
			\includegraphics[width=\linewidth, trim=2.0cm 2.0cm 2.0cm 2.0cm, clip]{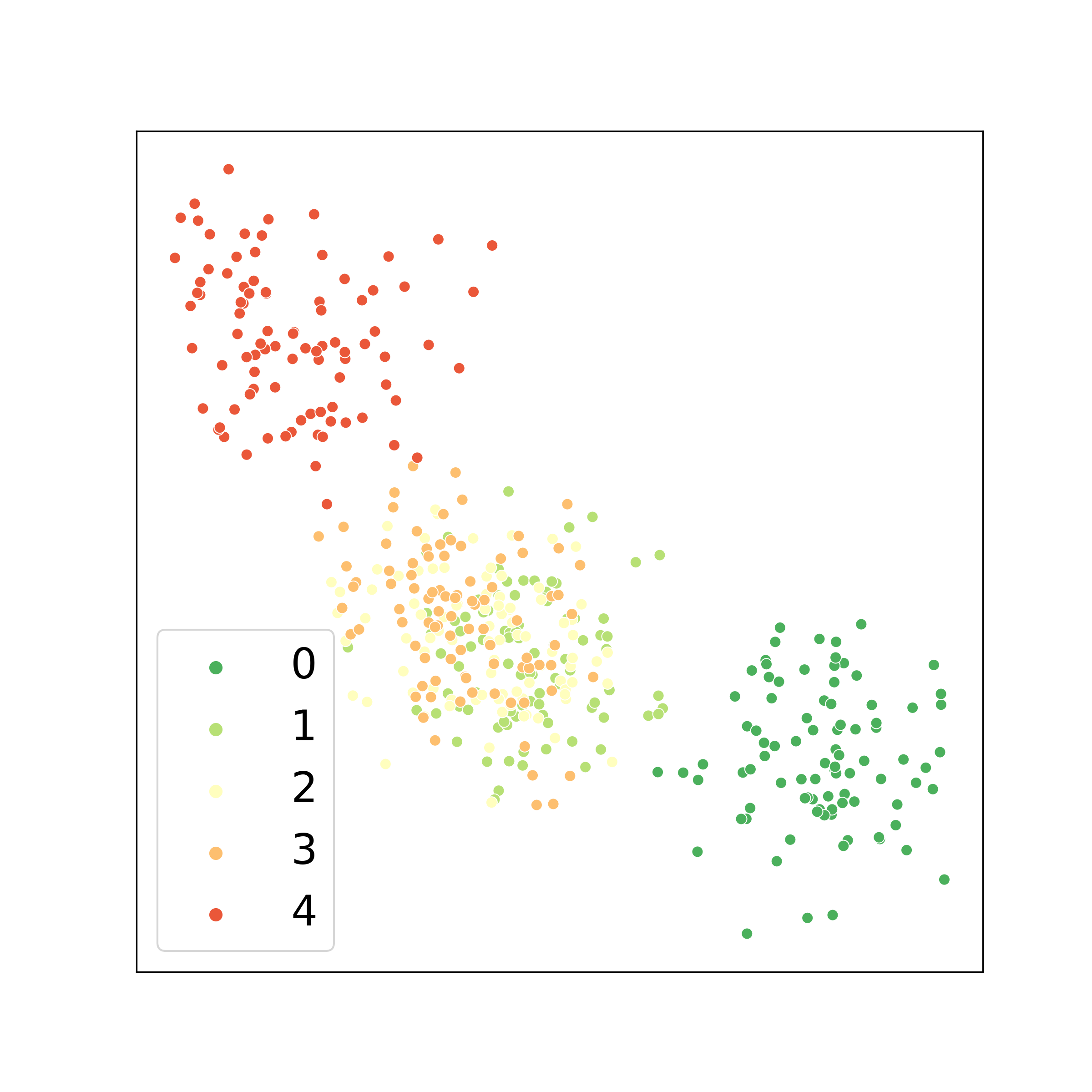}
			\caption{}
			\label{figPCAOne}
		\end{subfigure}
		\caption{\ac{acPCA} visualizations over all parameters coloured depending on (a) Feed speed, (b) Grit Size, and (c) a \ac{acPCA} visualization of only one sanding parameter configuration coloured depending on belt wear.}
		%\label{fig3}
	\end{minipage}
\end{figure}

The \ac{acPCA} visualization of \ac{acMFCC} features shows distinct clusters, unlike the Mel spectrograms, which show the best results in supervised classifications.
Without major loss of information, all recorded sanding processes, including all parameters, could be mapped to a meaningful 2D visualization using a \ac{acPCA} transformation.
Combined explained variance of the first and second \ac{acPCA} components are 88\%.
Fig. \ref{figPCASchnitt} shows the transformed data coloured depending on the sanding parameter Feed speed.
Fig. \ref{figPCAKoer} shows the same data coloured depending on Grit Size.
We also fitted a \ac{acPCA} for each sanding parameter configuration.
For some configurations, the transformations could visualize the abrasive belt wear meaningfully.
An example is shown in Fig. \ref{figPCAOne}.

\section{Conclusion}
\label{lblSectionConclusion}

We recommend using Mel Spectrograms and Decision Trees to determine abrasive belt wear of sanding machines in industrial processes due to their excellent accuracy and the possibility of using them on embedded devices in contrast to more complex classification models like neural networks.
Our results agree with \cite{GrindCurentSound}, which states that grinding is not a high-frequency problem.
Therefore, Mel Spectrograms and \ac{acMFCC} are more accurate for classifying abrasive belt wear than Spectrograms, \ac{acIMFCC} and \ac{acLFCC}.
Using \ac{acPCA} on \ac{acMFCC} features, we could visualize different belt wear levels, and the sanding parameters Feed speed and Grit Size meaningfully.
Since these visualizations look promising in two dimensions, further work includes unsupervised clustering methods.

\section*{Acknowledgment}

We would like to thank our industry partner Hans Weber Maschinenfabrik GmbH  for their cooperation and insight.

\clearpage

\bibliography{literature}
\bibliographystyle{ieeetr}

\end{document}